\documentclass[prl,twocolumn,superscriptaddress,showpacs,nobalancelastpage]{revtex4}

\usepackage{graphicx}

\begin{document}

\title{Non-destructive measurement of electron spins in a quantum dot}


\author{T. Meunier}
\affiliation{Kavli Institute of Nanoscience, Delft University of
Technology,\\
PO Box 5046, 2600 GA Delft, The Netherlands}
\author{I. T. Vink}
\affiliation{Kavli Institute of Nanoscience, Delft University of
Technology,\\
PO Box 5046, 2600 GA Delft, The Netherlands}
\author{L. H. Willems van Beveren}
\affiliation{Kavli Institute of Nanoscience, Delft University of
Technology,\\
PO Box 5046, 2600 GA Delft, The Netherlands}
\author{F. H. L. Koppens}
\affiliation{Kavli Institute of Nanoscience, Delft University of
Technology,\\
PO Box 5046, 2600 GA Delft, The Netherlands}
\author{H. P. Tranitz}
\affiliation{Institut f$\ddot{u}$r Angewandte und Experimentelle
Physik, Universit$\ddot{a}$t Regensburg, Regensburg, Germany}
\author{W. Wegscheider}
\affiliation{Institut f$\ddot{u}$r Angewandte und Experimentelle
Physik, Universit$\ddot{a}$t Regensburg, Regensburg, Germany}
\author{L. P. Kouwenhoven}
\affiliation{Kavli Institute of Nanoscience, Delft University of
Technology,\\
PO Box 5046, 2600 GA Delft, The Netherlands}
\author{L. M. K. Vandersypen}
\affiliation{Kavli Institute of Nanoscience, Delft University of
Technology,\\
PO Box 5046, 2600 GA Delft, The Netherlands}

\date{\today}

\begin{abstract}

We propose and implement a non-destructive measurement that
distinguishes between two-electron spin states in a quantum dot. In
contrast to earlier experiments with quantum dots, the spins are
left behind in the state corresponding to the measurement outcome.
By measuring the spin states twice within a time shorter than the
relaxation time, $T_1$, correlations between consecutive
measurements are observed. They disappear as the wait time between
measurements become comparable to $T_1$. The correlation between the
post-measurement state and the measurement outcome is measured
to be $\sim 90\%$ on average.

\end{abstract}


\maketitle

In standard quantum mechanics, repeated measurements of the same
observable produce the same outcome~\cite{Braginsky}. Read-out
schemes with this property are called \emph{non-destructive}. In
reality, a measurement of a quantum object often destroys the system
itself, in which case repeated measurements aren't possible. This is
the case, for instance, with conventional photon detectors. Even if
the quantum system itself is not destroyed by the measurement, its
state can be altered and a second measurement may give a different
result than the first measurement. An intrinsic property of
\emph{non-destructive} measurements is that the post-measurement
state corresponds to the measurement outcome. This characteristic is
of fundamental interest and also of practical relevance in the
context of quantum information processing. For instance,
non-destructive measurements can be used to quickly (re)initialize
selected qubits~\cite{DiVincenzo_crit}.

In quantum dots, non-destructive measurements of the charge state
have been implemented~\cite{FieldPRL,PettaPRL04}. For spin states in
quantum dots, however, all single-shot read-out schemes used so far
are destructive. Either the spin is always left in the  ground
state~\cite{NatureReadout}, or the number of electrons in the dot is
changed as a result of the measurement~\cite{RonaldPRL}. Here, we
present and implement a non-destructive, single-shot measurement
scheme that distinguishes two-electron singlet from triplet states
in a single quantum dot. We take advantage of the remarkably long
spin relaxation time,
$T_1$,~\cite{NatureReadout,RonaldPRL,FinleyNature}, to repeat the
measurement twice within $T_1$ and demonstrate experimentally that
the spin state after the read-out corresponds to the measurement
outcome.

Our measurement scheme is based on spin-to-charge conversion taking
advantage of a difference in tunnel rates between the dot and a
reservoir, depending on the spin state, as in Ref.~\cite{RonaldPRL}.
In the case of the singlet, both electrons are in the ground state
orbital whereas for the triplet state, one electron is in the first
excited orbital. The excited orbital has a stronger overlap with the
reservoir than the lowest orbital, causing the tunnel rate to and
from the triplet state, $\Gamma_T$, to be much larger than the
tunnel rate to and from the singlet state,
$\Gamma_S$~\cite{RonaldPRL}.

\begin{figure}[!t]
\includegraphics[width=3.4in]{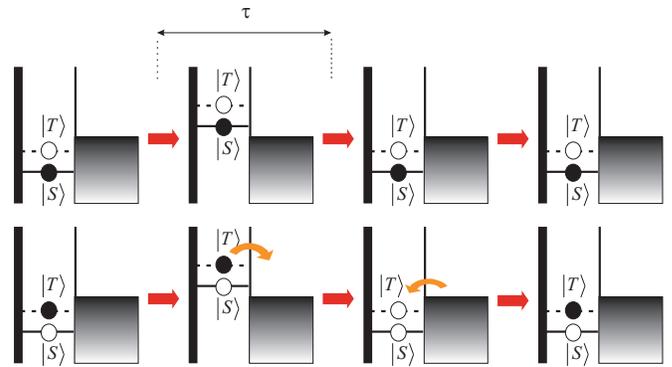}
\caption{Schematic of the quantum dot throughout the non-destructive
measurement scheme, for a singlet (Top) or triplet (Bottom) initial
state. Curved arrows indicate tunnel process. The spin state is the
same before and after the measurement.}
\label{Fig1}
\end{figure}

To implement the non-destructive measurement, we pulse the potential
of the dot so the electrochemical potential for both the singlet and
the triplet state lies above the Fermi energy for a short time
$\tau$ (see Fig.~\ref{Fig1}), fulfilling the relation $1/\Gamma_T
\ll \tau \ll 1/\Gamma_S$. In the experiment, $1/\Gamma_T \approx
5~\mu s$, $\tau=20~\mu s$, and $1/\Gamma_{S,out}=100~\mu s$ (for the
singlet, we observe the time to tunnel in is different from the time
to tunnel out: $1/\Gamma_{S,in}\approx 1000~\mu s$~\cite{note3}). If
the dot is in the singlet state, most of the time no electron
tunnels out during the entire pulse sequence since $\tau$ is small
in comparison with $1/\Gamma_S$, even though tunneling would be
energetically allowed. In the case of the triplet state, an electron
will tunnel off the dot after the pulse is applied, in a time
$1/\Gamma_T$ much smaller than $\tau$. In this case, an electron
tunnels back in after the pulse and, it will tunnel into the triplet
state with high probability since $\Gamma_T \gg \Gamma_S$.

The proposed read-out scheme is thus non-destructive in the sense
that the state after the measurement coincides with the measurement
result. The actual measurement takes place through the occurrence or
absence of the first tunnel process. For a superposition input
state, this is when the "projection" of the wave function would take
place. For a singlet initial state, the dot remains in the singlet
all along; for a triplet initial state, the dot is reinitialized
through a second tunnel event.

We point out that the proposed scheme is conceptually similar to the
measurement procedure used for trapped ions~\cite{WinelandRMP}. In
both systems, we can distinguish the two relevant states depending
on whether or not a transition is made through a third state (a
reservoir for the electron spin and a short-lived internal level for
the ion).

\begin{figure}[!t]
\includegraphics[width=3.4in]{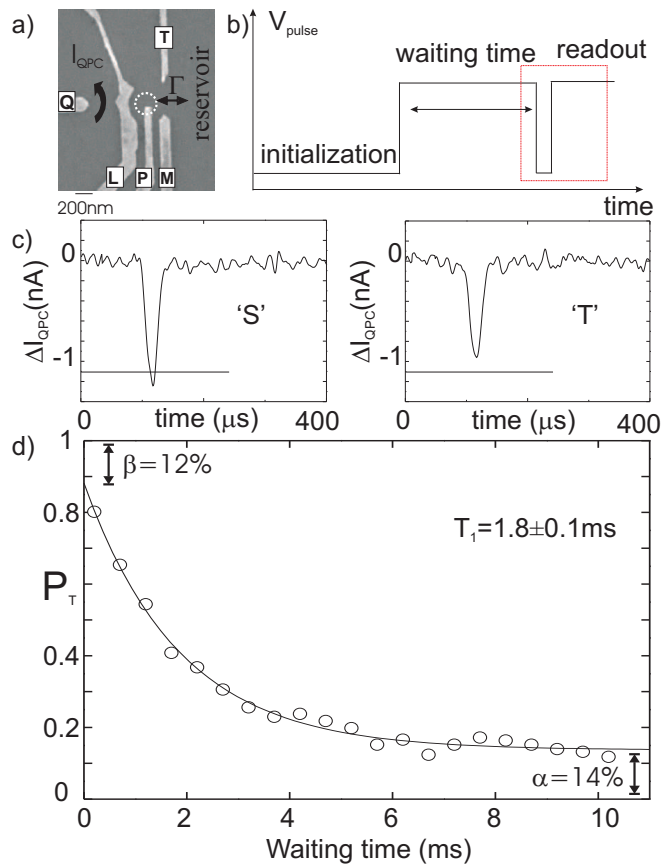}
\caption{(a) Scanning electron micrograph showing the sample design.
(b) Voltage pulses applied to gate $'P'$ for the relaxation
measurement. (c) Typical QPC response in a 400 $\mu$s interval
around the read-out pulse, for the case of singlet (left) and
triplet (right). The solid horizontal line indicates the position of
the threshold. (d) The probability for detecting a triplet state as
a function of the waiting time. Each point is an average over $500$
experiments. The solid line is an exponential fit to the data. The
measurement errors $\alpha$ and $\beta$ (see text) are indicated.}
\label{Fig2}
\end{figure}

We test this measurement concept with a quantum dot (white dotted
circle in Fig.~\ref{Fig2}(a)) and a quantum point contact (QPC)
defined in a two-dimensional electron gas (2DEG) with an electron
density of $1.3\cdot10^{15}$~m$^{-2}$, 90~nm below the surface of a
GaAs/AlGaAs heterostructure, by applying negative voltages to gates
$L$, $M$, $T$ and $Q$. Fast voltage pulses on gate $P$ are used to
rapidly change the electrochemical potential of the dot. All
measurements are performed at zero magnetic field. We tune the dot
to the few-electron regime~\cite{CiorgaPRB,JeroFewEl}, and
completely pinch off the tunnel barrier between gates $L$ and $T$,
so that the dot is only coupled to the reservoir on the
right~\cite{JeroAPL}. The conductance of the QPC is tuned to about
$e^2/h$, making it very sensitive to the number of electrons on the
dot~\cite{FieldPRL}. A voltage bias of 0.7~mV induces a current
through the QPC, $I_{QPC}$, of about 30~nA. Tunneling of an electron
on or off the dot gives steps in $I_{QPC}$ of 300~pA
\cite{LievenAPL,EnsslinAPL} and we observe them in the experiment
with a measurement bandwidth equal to 60~kHz.

First we demonstrate that the non-destructive measurement correctly
reads out the spin states. The experiment consists in reconstructing
a relaxation curve from the triplet to the singlet and comparing the
results with those obtained using destructive read-out
scheme~\cite{RonaldPRL}. The protocol is illustrated in
Fig.\ref{Fig2}(b). The starting point is a dot with one electron in
the ground state (initialization). In the second stage of the pulse,
the singlet and triplet electrochemical potentials are below the
Fermi energy and a second electron tunnels into the dot. Since
$\Gamma_T\gg\Gamma_S$, most likely a triplet state will be formed,
on a timescale of $1/\Gamma_T$. The non-destructive measurement
pulse is applied after a waiting time that we vary. Due to the
direct capacitive coupling of gate $P$ to the QPC channel, $\Delta
I_{QPC}$ follows the pulse shape (see Fig \ref{Fig2}(c)). The
precise amplitude of the QPC pulse response directly reflects the
charge state of the dot throughout the read-out pulse. If the two
electrons remain in the dot, the QPC signal goes below a predefined
threshold, and we conclude that the dot was in the singlet state
(outcome $'S'$, see Fig.~\ref{Fig2}(c), left). Otherwise, if one
electron tunnels out in a time shorter than the pulse response time,
the QPC pulse response stays above the threshold and we declare
that the dot was in the triplet state (outcome $'T'$, see
Fig.~\ref{Fig2}(c), right)~\cite{sigma}.

As expected, we observe an exponential decay of the triplet
population as a function of the waiting time, giving a relaxation
time, $T_1$, equal to $1.8\pm0.1$~ms. The measurement errors are
$\alpha=0.14$ and $\beta=0.12$, where $\alpha$ ($\beta$) is defined
as the probability for the measurement to return triplet (singlet)
if the actual state is singlet (triplet). We observe the same values
(within error bars) when using the known destructive read-out scheme
in this same measurement run. In both cases, measurement errors are
completely explained by the two different tunnel rates
\cite{RonaldPRL}. The resulting measurement fidelity,
$1-(\alpha+\beta)/2$, is 87\%. It is worth noticing that in this new
read-out scheme the measurement time, $t_{meas} \simeq \tau =
20~\mu$s, is much shorter than $T_1$ ($T_1/t_{meas}\simeq90$).

\begin{figure}[!t]
\includegraphics[width=3.4in]{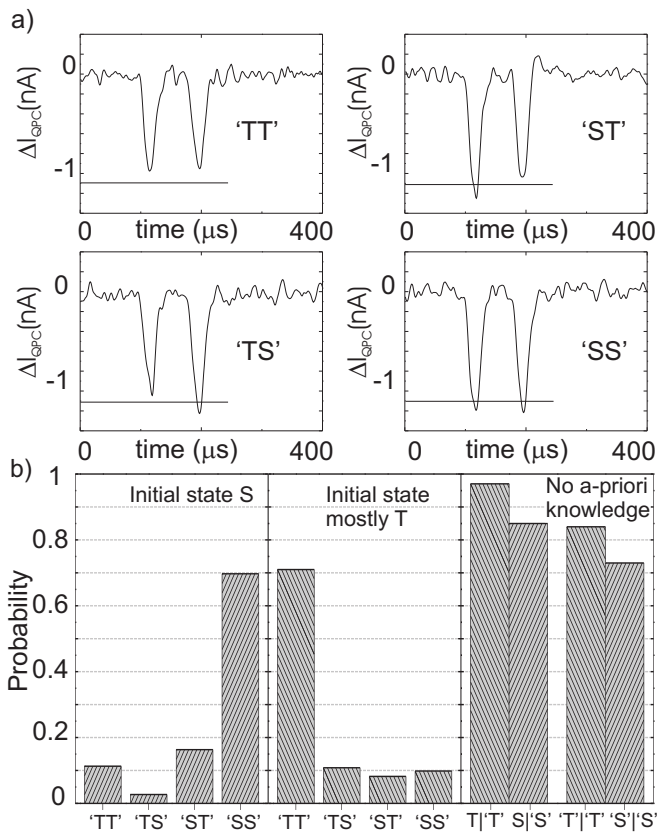}
\caption{(a) Typical QPC response for two consecutive measurements
in the case of $'SS'$, $'TT'$, $'ST'$ and $'TS'$. The threshold is
the same for the two non-destructive measurement pulses. The pulse
width is $20~\mu$s and the delay between the two measurement pulses
is $60~\mu$s. (b) The recorded probabilities for each of these four
events over 3000 runs, with the singlet (first graph) and mostly the
triplet (second graph) as the initial state. In the third graph,
conditional probabilities $P(T|'T')$ or $P(S|'S')$ that the state
after the first measurement corresponds to the outcome of the first
measurement and conditional probabilities $P('T'|'T')$ or
$P('S'|'S')$ that the second measurement gives the same outcome as
the first one are presented. They are extracted from the two
previous graphs and the known $\alpha$ and $\beta$ with no a-priori
knowledge of the initial state.} \label{Fig3}
\end{figure}

We next test if the measurement is non-destructive by studying the
correlations between the outcomes of two successive measurements. We
program a second read-out pulse $60~\mu$s after the end of the first
pulse and record the probability for each of the four combined
outcomes, $'SS'$, $'TT'$, $'ST'$, $'TS'$ (Fig.\ref{Fig3}). In order
to accurately characterize the measurement, we first do this with
singlet initial states (prepared by waiting $20$ ms for complete
relaxation), and then again with mostly triplet initial states
(prepared by letting the second electron tunnel in $200~\mu$s before
the first measurement~\cite{note1}). A clear correlation between
consecutive measurement outcomes is observed (Fig. 3(b)), both for
singlet and triplet initial states.

When we average over $S$ or $T$ initial states (i.e. when we have no
a-priori knowledge of the spin state), we find, from the correlation
data and the known values of $\alpha$ and $\beta$, an $85\%$
($73\%$) conditional probability for outcome $'T'$ ($'S'$) in the
second measurement given that the first measurement outcome was
$'T'$ ($'S'$)~\cite{note2}.

\begin{figure}[!t]
\includegraphics[width=3.4in]{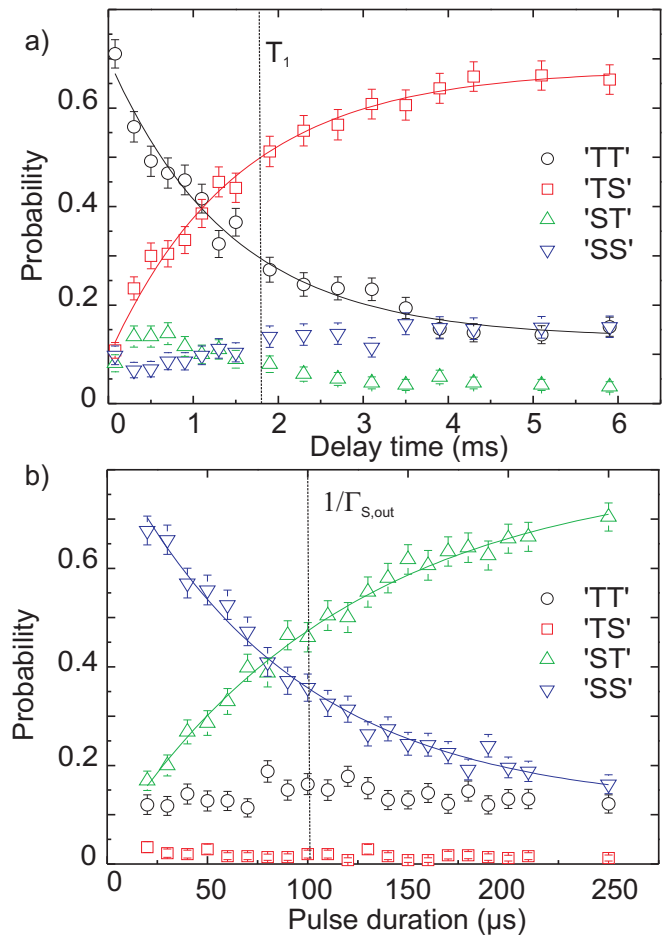}
\caption{The probabilities for the two consecutive measurement
outcomes as a function of (a) the  measurement delay and (b) the
measurement pulse duration. In solid line is represented the exponential
fit to the data.} \label{Fig4}
\end{figure}

The degree to which the scheme is non-destructive is quantified via
the probability for obtaining a $S$ or $T$ post measurement state
($60~\mu$s after the end of the first pulse) conditional on the
measurement outcome. From the correlation data and the known values
of $\alpha$ and $\beta$, we extract a $97\%$ ($84\%$) conditional
probability $P(T|'T')$ ($P(S|'S'))$, again assuming no a-priori
knowledge of the initial state~\cite{note2}.
For a triplet outcome, one electron tunneled out during the
measurement pulse, and another electron tunneled back in after the
pulse. A triplet state is formed with near certainty in this
reinitialization process (since $\Gamma_T/\Gamma_{S,in}\approx 200$),
but the triplet state can relax to the singlet during the $60~\mu$s
wait time between the two measurements. This occurs with a probability
$\gamma$ of $3\%$, which explains the observed conditional probability
$P(T|'T')$. The conditional probability
$P(S|'S')$ can be found as $1 - P(T,'S')/P('S')$. $P('S')$ is simply
$[(1-\alpha) + \beta]/2$ (averaged over $S$ and $T$ initial states).
There are two main contributions to $P(T,'S')$. First, for
$\beta=12\%$ of
the triplet initial states, both electrons remain on the dot. In
this case, a singlet outcome is declared but the post-measurement
state is almost always a triplet.
Second, for singlet initial states, a singlet
outcome is obtained with probability $1-\alpha=86\%$. For
$\sigma=5\%$ of those cases, one electron nevertheless tunneled
out and the post-measurement state is a triplet~\cite{note2}.

An attractive feature of non-destructive measurements is that it
allows one to study the time evolution between two successive
measurements. As a proof of principle, we let the spin evolve under
relaxation for a controlled time in between two measurements. The
singlet state is not affected by relaxation, so we initialize the
dot (mostly, as before) in the triplet state. In figure
\ref{Fig4}(a), the probabilities for the four possible outcomes are
recorded as a function of the waiting time. We notice that $'TT'$
and $'TS'$ respectively decay and increase exponentially, with a
time constant $1.5\pm0.3$~ms, within the error bars of the
relaxation time obtained from Fig.\ref{Fig2}(d).

Finally, we remark that the non-destructive nature of the
measurement relies on our ability to tune the dot in a regime where
$1/\Gamma_T\ll\tau\ll1/\Gamma_{S,out}$. If $\tau\gg1/\Gamma_{S,out},1/\Gamma_T$,
the measurement is destructive, because one electron will tunnel off
the dot during the read-out pulse irrespective of the state of the
dot. The information about the spin state is then lost after the
read-out and the post-measurement state will always be a triplet. We
can vary the duration of the pulse in order to make the transition
from non-destructive to destructive read-out. Here we initialize in
the singlet state, since for triplet initial states, the
post-measurement state doesn't change with $\tau$. Figure
\ref{Fig4}(b) summarizes the results. The four different curves
correspond to each combination of measurement outcomes as a function
of the duration of the pulse. As expected, the $'TS'$ and $'TT'$
statistics are steady, while the $'SS'$ and $'ST'$ probabilities
decay respectively increase exponentially with a time constant
$105\pm10~\mu s$, within the error bars of the evaluation of $1/\Gamma_{S,out}$.

In conclusion, we demonstrate our ability to implement a
non-destructive measurement scheme for distinguishing two-electron
singlet from triplet states in a single quantum dot. The spin system
is not strictly preserved throughout the entire measurement process.
In that respect, our scheme differs from a quantum non-demolition
(QND) measurement~\cite{Braginsky}. Nevertheless, repeated
measurements give the same results and the post-measurement state
corresponds to the measurement outcome. All the imperfections in the
correlations observed in the experiments are explained by the ratio
between the singlet and triplet tunnel rates, and the relaxation
rate from triplet to singlet. Other spin-dependent tunnel processes,
for instance as observed in double
dots~\cite{EngelPRL04,KoppensScience,JohnsonNature,PettaScience},
can be used for non-destructive read-out, possibly with even higher
fidelity.




\begin{acknowledgments}

We thank Ronald Hanson for useful discussions, Raymond Schouten and
Bram van der Enden for technical support and FOM, NWO and DARPA for
financial support.

\end{acknowledgments}

\begin{figure}
\centering
\includegraphics[width=3.4in]{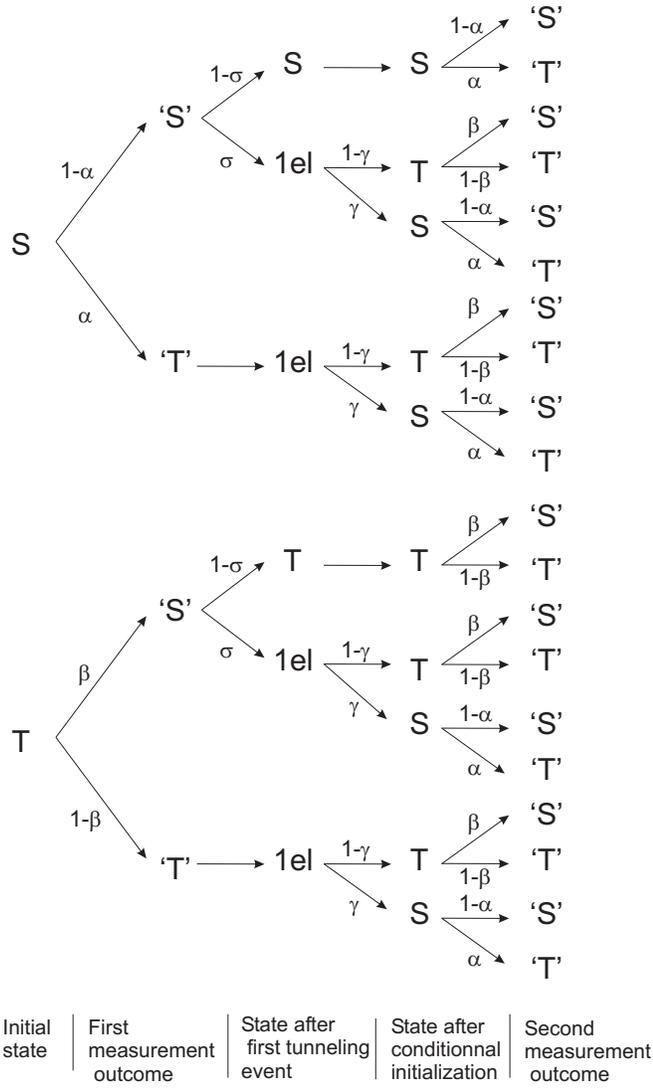}
\caption{Different events and their probabilities all along the
process of the two consecutive measurements for singlet S or triplet
T as an initial state. $\alpha$ and $\beta$ are defined as the
probability for the measurement to return respectively triplet and
singlet if the actual state is singlet and triplet. They are
obtained directly from the relaxation curve, giving
$\alpha = 14\%$ and $\beta = 12\%$. $\sigma$
is the probability for an electron to tunnel out even though the
QPC signal did go below the threshold and a $'S'$ outcome
has been declared (we assume that $\sigma$ is equal for singlet and
triplet initial states). Finally, $\gamma$ is the probability that,
after reinitializing to a two-electron state and subsequent
relaxation for $60~\mu$s, a singlet state is present in the dot.
In the experiment, $\sigma=5\%$ and $\gamma = 3\%$, determined
from the above probability tree, the known value of $\alpha$ and
$\beta$, and the correlation data.}
\end{figure}


\begin{references}
\bibitem{note3} A possible explanation could be that the pulse
not only shifts the dot potential but also distorts it, thereby
changing the orbitals.
\bibitem{sigma}
When the QPC signal went below the threshold and a $'S'$ outcome is
declared, there is still some probability, $\sigma$, that one electron
tunneled out during the pulse (after a time longer than the pulse
response time).
\end{references}



\end{document}